\documentclass[aps,prl,reprint,superscriptaddress, longbibliography]{revtex4-1}
\usepackage{H1}
\usepackage{amsmath}
\usepackage[pdftex,colorlinks=true]{hyperref}
\hypersetup{citecolor = blue,urlcolor=blue}
\usepackage[normalem]{ulem}
\usepackage{amssymb}
\usepackage{amsthm}
\usepackage{amsfonts}
\usepackage{bbm}
\usepackage{array}
\usepackage{braket}
\usepackage[dvipsnames]{xcolor}

\begin{document}
\title{Comment on ``Quantum Time Crystals from Hamiltonians with Long-Range Interactions"}

\author{Vedika Khemani}
\affiliation{Department of Physics, Stanford University, Stanford, CA 94305, USA}

\author{Roderich Moessner}
\affiliation{Max-Planck-Institut f\"{u}r Physik komplexer Systeme, 01187 Dresden, Germany}

\author{S. L. Sondhi}
\affiliation{Department of Physics, Princeton University, Princeton, NJ 08544}

\date\today

\begin{abstract}
In a recent paper (Phys. Rev. Lett. 123, 210602), Kozin and Kyriienko claim to realize  ``genuine'' ground state time crystals by studying models with long-ranged and infinite-body interactions. Here we point out that their models are doubly problematic: they are unrealizable {\it and} they violate well established principles for defining phases of matter. Indeed with infinite body operators allowed, almost all  quantum systems are time crystals. In addition, one of their models is highly unstable and another amounts to isolating, via fine tuning, a single degree of freedom in a many body system--allowing for this elevates the pendulum of Galileo and Huygens to a genuine time crystal. 
\end{abstract}

\maketitle

In a paper with the delightful title ``Exact questions to some interesting  answers  in  many  body  physics", Arovas and Girvin noted 
that \emph{``it is always possible to choose a Hamiltonian ${\cal H}=-|\Psi\rangle\langle\Psi|$ which renders [any {\it given} wavefunction] $|\Psi\rangle$ its exact nondegenerate ground state. Such a Hamiltonian, however, will generally involve interactions among arbitrarily large numbers of particles and over arbitrarily long distances, and will not be the sort of model that captivates the interest of one's colleagues.}"~\cite{Arovas1992}. Kozin and Kyriienko (KK) \cite{Kozin} have done what Arovas and Girvin ruled out, and have written down models of infinite range and with infinite-body interactions. This comment  elaborates on this quotation.

The first point is that general Hamiltonians with $N$-body {\it instantaneous} interactions for systems of $N$ particles are simply not realizable beyond small values of $N$. The interactions that we take to be fundamental in the laboratory 
are all few body.  Integrating out other degrees of freedom will result in multi-particle interactions with non-trivial retardation. This issue may be elided when these multi-body terms are small but not when the dominant term is supposed to be an $O(N)$-body interaction. 

Even if we put that aside, the second point is that the entire theoretical exercise of  defining and finding phases of matter is predicated on a degree of spatial locality and becomes problematic when this is abandoned. Allowing such constructions would have dramatically short-circuited the quest for many novel quantum phases of matter such as spin-liquids. And once one admits Hamiltonian operators with $N$-body interactions there is no reason not to admit $N$-body observables, at which point one is simply talking about a zero dimensional system with an enormous Hilbert space. Indeed, if $N$-body observables are admitted, {\it all} quantum systems exhibit oscillations in $\langle  m|O(t)O| m\rangle$ out to arbitrarily late times, for all Hamiltonian eigenstates $|m \rangle$ and for an infinite set of choices $O_{nm}=|n \rangle\langle m| + {\rm h.c}$ \cite{TTSBRep}, and thus every system is a time crystal at all energies, not just in its ground state.

We would like to take this opportunity to point out some further special features of the models that KK do present.
KK search for model Hamiltonians for which the $|GS\rangle$ satisfies the property that  $M_z|GS\rangle \propto |ES\rangle$, where $|ES\rangle$ is some excited eigenstate that is orthogonal to the ground state and non-degenerate with it, and $M_z$ is the total magnetization $M_z = \sum_i \sigma_i^z$ for spin 1/2 Pauli operators $\sigma^\alpha_i$. If this is satisfied, then $C(t) \equiv \lim_{V\rightarrow \infty} \frac{1}{V^2} \langle  GS|M_z(t)M(0)| GS\rangle $, where $V$ is the system volume, oscillates with a frequency set by the energy difference betweeen $|GS\rangle$ and $|ES\rangle$. Such oscillations, first discussed by Watanabe and Oshikawa \cite{Oshikawa15}, are one of the diagnostics of time crystal behavior \cite{briefhistory}.

The first model Hamiltonian in Ref.~\cite{Kozin}, Eq.~5, is precisely a Girvin-Arovas projector, $\mathcal{H}=-|G_+\rangle\langle G_+|$ where $|G_+\rangle$ is the GHZ Schr\"odinger cat state, $|G_{\pm}\rangle = \frac{1}{\sqrt 2}\left[ |\uparrow \uparrow \cdots \uparrow\rangle \pm |\downarrow \downarrow \cdots \downarrow\rangle\right]$. This has one ground state with energy $E=0$ and $2^N -1$ degenerate excited states with energy $E=1$. This model is obviously highly non-local and, further, given the high degeneracy of the excited state manifold,  the oscillations are not stable to the inclusion of generic perturbations such as Heisenberg exchange or a random onsite field~\footnote{Note also the absence of a factor of $N$ in order to obtain an extensive ground state energy. In its absence, an extensive perturbation, e.g. $M_z$ itself, becomes a singular perturbation and can completely change the nature of the ground state. However, in its presence, the oscillations that KK present are pathological in that they become infinitely rapid in the infinite system limit.}.

The second model Hamiltonian, Eq.~9 in Ref~\cite{Kozin}, can be exactly rewritten exclusively in terms of global operators, as ${\cal H} = \frac{J}{2N(N-1)} P_x \left(
M_z^2 - N
\right)$, where $P_x=\prod_{i=1}^N \sigma^x$ is a {\it global} spin-flip operator, and $M_z$ is the concomitant {\it total} $z$ component of the magnetization. The system thus effectively ceases to be a many-body system, and is described instead only by one global degree of freedom, whose Hilbert space scales linearly with $N$. 

However, for phase transitions in general, and time crystals in particular, the many-body nature of the system is not incidental but constitutive. To wit, the arguably simplest few-body system, a two-level Rabi oscillator, exhibits time-dependence in its ground state; hence, a genuinely non-trivial incarnation of a TC necessarily requires one to work with a \emph{many-body} system in the infinite volume limit. 
Indeed, back in 1960, Brout noted that ``it is always remarked that in
the limit that the exchange potential becomes very
long range [...], the molecular
field theory is recovered as a limit"\cite{Brout60} and there certainly has been no shortage of mean-field time crystals \cite{briefhistory}. 

Indeed, if one decides to alter the rules of the game to allow unphysical schemes to remove unwanted physical processes, we could straightforwardly endow a simple pendulum with the capacity to exhibit perennial periodic oscillations. In reality, the oscillations in a \emph{many-body} pendulum degrade, even in the absence of friction, by redistributing energy from the centre-of-mass degree of freedom involved in the oscillations to the random thermal motion of numerous internal (vibrational etc.) degrees of freedom of the pendulum \cite{briefhistory}. By endowing these other modes with infinite stiffness, or otherwise projecting them out, their excitation can be entirely suppressed. This scheme thus extends the realm of time crystals to include the pendulum clock itself. It would thus seem unfair not to credit Galileo and Huygens with the original discovery of a ``genuine time crystal" \cite{wiki:Pendulum}, as all they were missing was the unphysical part of the argument establishing time-translational symmetry-breaking.


Finally the third Hamiltonian in Ref.~\cite{Kozin}, Eqs. 10 and 11, also involves global spin operations, simultaneously acting on half the spins in a chain, \emph{i.e.} still an $O(N)$ interaction like the previous models. 
Indeed, all the constructions presented in Ref.~\cite{Kozin} \emph{constitutively} need \emph{at least} $N/2$ body interactions~\cite{Kozin}. There is, of course, a long history in physics of devising and using tractable models to gain insight -- but the resulting insights should not crucially depend on the non-physical aspects of these models.

Finally, once unphysical Hamiltonians are fair game, the above constructions were not really needed: 
Floquet unitaries that give rise to stable discrete time crystals (and arise from local time-periodic Hamiltonians, $H(t+T)=H(t)$),
are already known from the discovery of the phase~\cite{khemani_phase_2016, Q}. The Floquet unitary is defined as the time-evolution operator over one period, $U_F = \mathcal{T} e^{-i\int_0^T dt H(t)}$; the corresponding static Floquet Hamiltonians, $H_F$, defined via, $U_F \equiv e^{-iH_F T}$, give rise, by construction, to identical physics. However, these are known to be generically unphysical as static Hamiltonians due to their non-locality~\cite{BukovPolkovnikov2015}---which is precisely why \emph{new Floquet phases} exist.

\bibliography{comment}

\end{document}